\documentclass[journal]{IEEEtran}
\IEEEoverridecommandlockouts
\usepackage{times}
\usepackage{cite}
\usepackage[dvips]{color}
\usepackage{epsf}
\usepackage{epsfig}
\usepackage{graphicx}
\usepackage{graphics}
\usepackage{epstopdf}
\usepackage[small]{caption}
\usepackage[bottom]{footmisc}
\usepackage{amsmath}
\usepackage{amssymb}
\usepackage{amsthm}
\usepackage{amsxtra}
\usepackage{algorithm}
\usepackage{algorithmic} 	
\usepackage{enumerate}
\usepackage{multirow}
\usepackage{enumerate}
\usepackage{amssymb}
\usepackage{lipsum}
\usepackage{setspace}
\usepackage{multirow}
\usepackage{tikz}
\usepackage{setspace}
\usepackage{subfigure}
\usepackage{longtable}

\usepackage{rotating} 
\usepackage{graphicx} 

\newcommand{\Rmnum}[1]{\expandafter\@slowromancap\romannumeral #1@}

\makeatother
\begin{document}
\title{Cost Minimization of Charging Stations with Photovoltaics: An Approach with EV Classification}
\author{Wayes~Tushar,~\IEEEmembership{Member,~IEEE,}
        Chau~Yuen,~\IEEEmembership{Senior Member,~IEEE,}
        Shisheng~Huang,~\IEEEmembership{Member,~IEEE,}
        David~B.~Smith,~\IEEEmembership{Member,~IEEE,}
        and
        H.~Vincent~Poor,~\IEEEmembership{Fellow,~IEEE}
\thanks{W. Tushar and C. Yuen are with the Singapore University of Technology and Design (SUTD), 8 Somapah Road, Singapore 487372. (Email: \{wayes\_tushar,~yuenchau\}@sutd.edu.sg).}
\thanks{S. Huang is with the Operations Research Unit of the Ministry of Home Affairs, Singapore.~(Email:~shisheng\_huang@gmail.com).}
\thanks{D. B. Smith is with National ICT Australia (NICTA), ACT 2601, Australia. He also is an adjunct fellow with the Australian National University (ANU), ACT, Australia.~(Email: david.smith@nicta.com.au).}
\thanks{H. V. Poor is with the School of Engineering and Applied Science, Princeton University, Princeton, NJ 08544, USA~(Email:~poor@princeton.edu).}
\thanks{This work is supported by the Singapore University of Technology and Design (SUTD) through Energy Innovation Research Program (EIRP) Singapore NRF2012EWT-EIRP002-045.}
\thanks{D. B. Smith's work is supported by NICTA, which is funded by the Australian Government through the Department of Communications and the Australian Research Council.}
}
\IEEEoverridecommandlockouts
\maketitle
\begin{abstract}
This paper proposes a novel electric vehicle (EV) classification scheme for a photovoltaic (PV) powered EV charging station (CS) that reduces the effect of intermittency of electricity supply as well as reducing the cost of energy trading of the CS. Since not all EV drivers would like to be environmentally friendly, all vehicles in the CS are divided into three categories: 1) premium, 2) conservative, and 3) green, according to their charging behavior. Premium and conservative EVs are considered to be interested only in charging their batteries, with noticeably higher rate of charging for premium EVs. Green vehicles are more environmentally friendly, and thus assist the CS to reduce its cost of energy trading by allowing the CS to use their batteries as distributed storage. A different charging scheme is proposed for each type of EV, which is adopted by the CS to encourage more EVs to be green. A basic mixed integer programming (MIP) technique is used to facilitate the proposed classification scheme. It is shown that the uncertainty in PV generation can be effectively compensated, along with minimization of total cost of energy trading to the CS, by consolidating more green EVs. Real solar and pricing data are used for performance analysis of the system. It is demonstrated that the total cost to the CS reduces considerably as the percentage of green vehicles increases, and also that the contributions of green EVs in winter are greater than those in summer.
\end{abstract}

\section*{Nomenclature}
\addcontentsline{toc}{section}{Nomenclature}
\begin{IEEEdescription}[\IEEEsetlabelwidth{$V_1,~~V_2,$}]
\item[$T$] Total number of considered time slots.
\item[$t$] Time slot index.
\item[$I_\text{light}(t)$] Intensity of solar radiation at $t$.
\item[$K$] Total number of solar panels.
\item[$A$] Area of each panel.
\item[$\kappa$] Efficiency of each panel.
\item[$N$] Total number of Electric Vehicles (EVs) in the CS.
\item[$n$] Index of each EV.
\item[$t_{n,a}$] Arrival time of EV $n$ to the CS.
\item[$t_{n,l}$] Leaving time of EV $n$ from the CS.
\item[$t_{n,l}^\text{early}$] Early leaving time of green EV $n$ in case of emergency.
\item[$b_n$] Battery capacity of EV $n$.
\item[$s_n(t)$] State of charge (SOC) of the battery of EV $n$ at $t$.
\item[$s_{n,r}$] Target SOC of EV $n$ at $t_{n,l}$.
\item [$s_{n,f}^\text{min}$] Minimum SOC required by green EV $n$ before leaving the CS during any emergency.
\item [$s_{n,\text{min}}$] Minimum SOC requirement by the battery of any EV $n$.
\item[$\mathcal{N}$] The set of EVs in the CS.
\item[$\mathcal{N}_s$] The set of premium EVs.
\item[$\mathcal{N}_c$] The set of conservative EVs.
\item[$\mathcal{N}_f$] The set of green EVs.
\item[$e_\text{gen}(t)$] Generated solar energy from roof-top solar panels at $t$.
\item[$e_\text{pv}$] Generated solar energy from a single PV module.
\item[$e_{g,\text{out}}(t)$] Amount of electricity that the CS buys from the grid at $t$.
\item[$e_{g,\text{in}}(t)$] Amount of electricity that the CS sells to the grid at $t$.
\item[$e_{n,c}^s$] The charging rate for premium EVs.
\item[$e_{n,c}^{cn}$] The charging rate for conservative EVs.
\item[$e_{n,c}^{f}$] The charging rate for green EVs.
\item[$e_{n,d}^{f}$] The discharging rate for green EVs.
\item [$e_{n,\text{max}}$] The maximum allowable charging rate for any EV $n$.
\item [$p_{n,c}^s(t)$] Charging price per unit of energy for premium EV $n$ at $t$.
\item [$p_{n,c}^{cn}(t)$] Charging price per unit of energy for conservative EV $n$ at $t$.
\item [$p_{n,c}^{f}(t)$] Charging price per unit of energy for green EV $n$ at $t$.
\item [$p_{n,d}^{f}(t)$] Discharging price per unit of energy for green EV $n$ at $t$.
\item[$p_{g,\text{in}}(t)$] Buy price of electricity by the grid from the CS at $t$.
\item[$p_{g,\text{out}}(t)$] Sell price of electricity by the grid to the CS at $t$.
\item[$\rho$] The mark up price that the CS adds to $p_{g,\text{out}}(t)$ for selling the energy to the EVs.
\item [$\eta$] A predefined parameter set by the CS to determine $p_{n,c}^{f}(t)$.
\item [$\epsilon$] A predefined parameter set by the CS to determine $p_{n,d}^{f}(t)$.
\item [$\gamma$] A predefined parameter set by the CS to determine $p_{n,c}^{cn}(t)$.
\item [$\mu$] Battery efficiency.
\item [$y_n(t)$] Binary variable associated with each green EV $n$ for determining the direction of flow of energy at $t$.
\item [$x(t)$] Binary variable associated with the grid for determining the direction of flow of energy $t$.
\end{IEEEdescription}
\section{Introduction}\label{sec:introduction}
\IEEEPARstart{T}{he} electricity and transportation industries contribute approximately 64\% of the total global carbon dioxide (CO$_2$) production \cite{Labatt-Book:2007}, which is an increasing concern due to its detrimental impact on the environment. As a promising solution to this problem, adoption of electric vehicles (EVs) and renewable energy sources can considerably reduce emissions from the transportation and power sectors respectively~\cite{Yousuf-saber-TIE:2011}. However, to procure the most environmental and economical benefits, EVs and renewable energy sources need to be deployed together in a smart grid. For example, if a large fleet of EVs are connected to the grid and their required electricity is entirely produced by a coal-fired power plant, the charging of EVs will still produce a significant amount of CO$_2$~\cite{Masuch-C-IEVC:2012}.

One possible strategy to reduce CO$_2$ emissions is to use renewable energy as a complete or partial source of power for EV charging stations (CSs)~\cite{Markel-CVPPC:2009}. However, unregulated charging of EVs poses a significant risk to the smart grid system in terms of overloading network components~\cite{Shisheng-EnergyPolicy:2012}, power losses, and voltage deviations~\cite{Clement-Nyns-TPS:2010}. Furthermore, deploying renewable sources is challenging due to their variable electricity production, and consequently the uncertainty of supply of continuous power. In this context, significant research efforts have been made in recent years to explore suitable solutions for deploying renewables~\cite{Yousuf-saber-TIE:2011}, and of EV charging in both vehicle-to-grid (V2G) and grid-to-vehicle (G2V) settings~\cite{Shisheng-WindEnergy:2012}.

Most EV research to date, as we will see in the next section, has mainly focused on the interconnection of energy storage of vehicles and the grid with goals to exploit the environmental and economic benefits of EVs. There has also been a growing interest in exploring the potential of solar energy as a complete or partial source of power for a CS~\cite{Birnie-JPS:2013}. However, very little has been done regarding the optimal management of energy of a CS with photovoltaics with a view to minimize its operational cost of energy trading with connected EVs and the main electricity grid of the system. As a consequence, there needs to be solutions that can capture the intermittency of PV availability, random changes of vehicle traffic in a CS, and the variability of electricity prices across different times of the day, and that also minimizes the operational cost to a CS. We stress that a considerable number of different energy trading techniques, e.g., based on constraint optimization and game theory, are available in the literature~\cite{Fang-J-CST:2012}. However, most of these techniques require sophisticated computational intelligence only available to a future smart grid. Nonetheless, it is also important to devise solutions that can be developed with current system structures, with minor modification, if required, as well as solutions that are capable of efficiently managing energy trading within a smart grid system. Besides, the solutions also need be simple enough to be integrated with other existing optimization schemes for further performance improvement.

To that end, we propose a novel EV classification scheme in this paper. Based on the users' environmental friendliness, EVs are categorized into three types: premium, conservative and green. Premium EVs have the highest possible charging rate, and this ensure that the batteries of EVs are charged to the maximum possible states \emph{whenever} the EVs leave the CS. Conservative EVs have their battery charged to their desired state when the EVs leave the CS at their pre-defined time slot. Meanwhile, green EVs are similar to conservative EVs except that their batteries could be discharged, and hence if any EV leaves earlier than the pre-defined time slot, the battery state could be lower. We design different incentive schemes for each of these three types of EVs in terms of charging rates and charging prices. The composition of different charging rates and charging prices for different type of EVs are based on the real-time price offered by the grid at different times of the day, e.g, we choose the real time price from \cite{PJM:2013}. 

While designing different types of EVs it is considered that premium and conservative vehicles only prefer to charge their batteries and do not allow the CS to discharge. However, their difference between each other stems from their choice of different rates of charging, and the way they are debited by the CS during charging. Green EVs are assumed to be more environmental friendly, and thus allow the CS to use their batteries in compensating the uncertainty of PV production by accumulating and releasing energy at different times of their stay during energy trading. Therefore, the total cost to the CS reduces considerably as the relative number of green EVs increase in the system. However, during times of scarcity of energy from solar panels and green EVs, the CS can also buy electricity from the grid at a higher price. Therefore, on one hand, the CS can control the charging and discharging process of each EV and minimize its total cost of energy trading by providing more incentives to the connected EVs to be green. On the other hand, each EV can decide, based on its preference, on which type of vehicle it wants to be during its charging period at the CS. To study the effectiveness of the proposed classification and incentives scheme, we adopt a basic mixed-integer programming (MIP) approach, in which the objective function and constraints are designed to facilitate the proposed classification of EVs. We further show the validity of the scheme and its beneficial properties via extensive simulation.

In summary, the main contributions of this work are: 1) a novel EV classification scheme leveraged by users environmental friendliness and expected charging rate, in which EVs are categorized into premium, conservative and green; 2) design of different incentive schemes for each of these three types of EVs in terms of charging rate and charging/discharging price, where the formation of different charging rates and charging/discharging prices are based on real-time prices offered by the grid at different times of the day; 3) modeling of a scheduler based on the basic MIP approach, which is suitable to facilitate the proposed EV classification scheme with relevant objectives and constraints; and 4) finally, extensive numerical results on the performance of the scheme to establish the validity and illustrate the beneficial properties of the proposed EV classification scheme.

The rest of the paper is organized as follows: We discuss the state-of-the art of EV research in conjunction with renewables in Section~\ref{sec:literature-review}, which is followed by the demonstration of the system model and the proposed EV classification in Section~\ref{sec:system-model}. In Section~\ref{sec:problem-formulation}, we discuss the basic MIP approach, and design the objective function and constraints that facilitate the proposed classification scheme. Numerical experiments using real-data are conducted in Section~\ref{sec:numerical-simulation}, and finally we draw some concluding remarks in Section~\ref{sec:conclusion}.

\section{State-of-The Art}\label{sec:literature-review}
To address different challenges related to integrate EVs into smart grid, a major focus of recent literature has been devoted to model coordinated charging schemes from different view points of the grid system such as: to minimize the power losses and voltage deviation \cite{Clement-Nyns-TPS:2010};  to maximize the delivered amount of energy without exceeding network capacity limits~\cite{Richardson-TPS:2012}; to reduce the generation and other associated costs~\cite{Deilami-TSG:2011,Z-Ma-TCST:2013}; and to design the settlement between different energy entities considering their individual interests~\cite{Junjie-Hu-TSG:2014}. Further, intelligent scheduling of EVs, both as loads and sources, are investigated in \cite{Yousuf-saber-TIE:2011} in order to evolve a sustainable integrated electricity and transportation infrastructure, whereby scheduling is studied in \cite{Z-Ma-TCST:2013} and \cite{Yifeng-He-TSG:2012} for shaping the over-night demand curve, and to minimize the charging and discharging cost. In \cite{Lan-J-NCA:2013}, the authors studied an optimal control of EV charging schedule by characterizing a non-linear battery model. Further, a particle swarm optimization (PSO) based charging scheduling is discussed under dynamic electricity price in \cite{Jian-An-C-ICNSC:2013} and for the case of distributed generator failure in \cite{Kang-J-TASE:2013}. Other decentralized EV charging and energy management schemes can be found in \cite{Gan-TPS:2013,Rezaei-TSG:2014,Wayes-J-TSG:2012,Tushar-TIE:2014,Naveed_TSG:2015,LiuYi_TIE:2015,LiuYi_TSTSP:2014,Naveed-Energies:2013} and \cite{Hui-Liu-TPS:2013}.

EVs have also been extensively used to provide ancillary services such as providing energy storage to smooth the intermittency of renewables and spinning reserve \cite{Ota-TSG:2012} and to enable frequency regulation by efficient use of distributed storages  \cite{Sekyung-Han-TSG:2010}. However, as discussed in Section~\ref{sec:introduction}, EVs cannot provide environmental friendly and cost-effective solutions alone unless they are powered by environmental friendly energy sources \cite{Singh-Energy:2013} such as wind and solar. The studies on this topic have primarily focussed on the integration of EVs with wind energy farms \cite{Nina_Juul-Energy:2011}. Nonetheless, there has also been a number of studies such as the literature surveyed in \cite{Fang-J-CST:2012}, which discuss the potential of EVs in assisting solar integration in smart grid. However, the application of solar as an \emph{energy source for the CS} has not yet been well exploited \cite{Richardson-RSER:2013}.

To this end, there has been a growing interest in exploring the potential of solar energy as a complete or partial source of power for CS recently. In \cite{Birnie-JPS:2013}, for instance, a solar built parking lot is proposed for charging commuter vehicles, which can effectively meet the driving needs of summer in New Jersey. Suitable sizes of solar panels for charging a plug-in hybrid electric vehicle (PHEV) are investigated in \cite{X-Li-PES:2009}, where the authors show the different required area of solar panels for summer and winter to exclusively charge the EV via photovoltaics. Large scale development of parking lot solar car chargers  is analyzed in \cite{Neumann-PPRA:2012} and it is demonstrated that $14-50\%$ of the city's passenger transportation energy demand could be provided through solar energy. Further, the technical feasibility of charging EVs through on-site solar electricity and their use to improve vehicle efficiency are studied in \cite{Gibson-JPS:2010,Kelly-JPS:2010,Giannouli-SolarEnergy:2012}.

However, as can be seen from the above discussion, the optimal management of energy for a PV powered CS in order to minimize its cost of operation has not yet received significant attention. Particularly, there is a need to devise solutions that will minimize the operational cost to a CS that arises from managing its electricity across different time slots of the day. Furthermore, the solutions need to be easily implementable and should be captured through existing techniques. To this end, we propose a suitable system model that will capture these characteristics and design an energy trading scheme for a PV powered solar station through a novel vehicle classification scheme in the next section.

\section{Model and EV Classification}\label{sec:system-model}
\begin{figure}[t]
\centering
\includegraphics[width=\columnwidth]{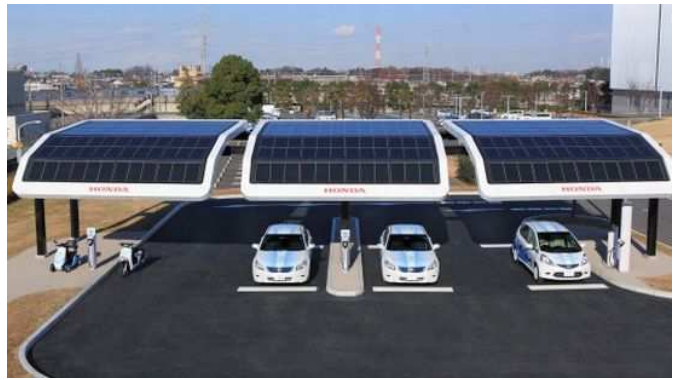}
\caption{An EV charging station with roof-top solar panels ~\cite{SolarEVStation}.} \label{figure:EV-solat_station}
\end{figure}

Let us consider a smart grid system that consists of a single CS and a number of EVs that are connected to it. The CS is equipped with roof-top solar panels, such as the Baldwin Parking Lot in Downtown Westport, Connecticut, USA~\cite{SolarEVStation} as shown in Fig.~\ref{figure:EV-solat_station}, that can produce energy from sunlight, and use the energy to charge the connected vehicles. The CS is also connected to the main grid so that it can buy energy from the grid in times of energy deficiency. In order to increase the use of renewable energy in the system, the CS can also discharge the batteries of connected EVs, if possible, and thus, keeps the consumption of energy from the grid to a minimum. Nevertheless, if the cost-benefit analysis is favorable, the CS can also sell its excess energy (if there is any) to the main grid.

\begin{figure}[t]
\centering
\includegraphics[width=\columnwidth]{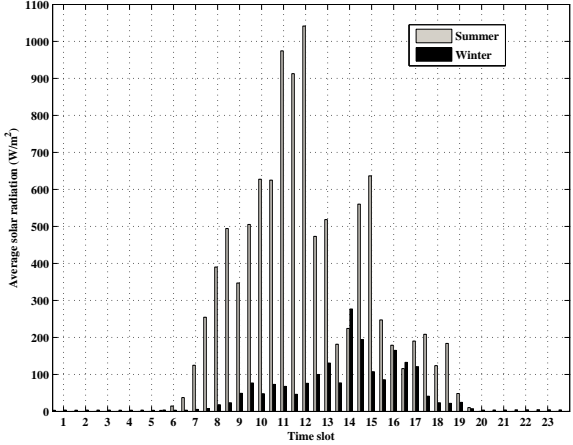}
\caption{Solar radiation data from Australian National University.} \label{figure:solar-data}
\end{figure}
We assume a per time-slot based energy trading scheme for the CS. The total scheduling time is assumed to have a duration of $11$ hours, i.e., $7$ am to $6$ pm considering the availability of the sun above the horizon in both summer and winter. The total time is segmented into $T = 22$ time slots, where each time slot $t$ has a duration of $30$ minutes~\cite{Anderson-W-EIS:2010}. It is important to note that the electricity market usually operates on an hourly/ half hourly basis and hence it is reasonable to divide the complete time domain into suitable time slots~\cite{Jin-J-TVT:2013}. The generation of solar energy per time slot $e_\text{gen}(t)$ by the roof-top solar panel is calculated based on the solar radiation data $I_\text{light}(t)$ measured at the Center for Sustainable Energy Systems at Australian National University~\cite{ANU-Solar-data:2013}. For instance, the solar irradiance data for a typical day in summer and winter are shown in Fig.~\ref{figure:solar-data}.  We presume that the CS requires $K$ solar panels to cover its entire parking space. To that end, first we note that the output energy $e_\text{gen}$ of a photovoltaic (PV) array is the product of the total energy $e_\text{pv}$ produced by a single PV module\footnote{A PV module/panel represents the fundamental power conversion unit of a PV system~\cite{Zhou-J-AE:2010}.} and the total number of solar panels $K$ in the entire array~\cite{Zhou-J-AE:2010}. Secondly, based on the efficiency of a PV solar panel $\kappa$, the output of a single solar panel $e_\text{pv}$ is related to the solar intensity $I_\text{light}$ through the following relationship~\cite{PV-basic}:
\begin{equation}
e_\text{pv} = \kappa\times A\times I_\text{light},
\label{eqn:energy-basic}
\end{equation}
where $A$ is the area of each solar panel. Therefore, if the solar intensity at time slot $t$ is $I_\text{light}(t)$, the solar generation $e_\text{gen}(t)$ from the roof-top solar panels can be considered to be
\begin{equation}
e_\text{gen}(t) = \kappa\times I_\text{light}(t) \times A\times K.
\label{eqn:solar-output}
\end{equation}
We assume that the CS uses the generated energy $e_\text{gen}(t)$ to charge the connected vehicles at time $t$. However, the CS can also buy $e_{g,\text{out}} (t)$ from, or sell $e_{g,\text{in}}(t)$ to, the main grid if such trading is convenient.

To this end, let us consider that $\mathcal{N}$ is the set of total vehicles in the CS, where $|\mathcal{N}|=N$, and each vehicle $n\in\mathcal{N}$ can arrive at and leave from the CS at any time during $T$. Each vehicle has a different arrival and leaving time, $t_{n,a}\in T$ and $t_{n,l}\in T$ respectively. The  battery capacity of each vehicle is $b_n$ and the state of charge (SOC) of the battery is $s_n(t),~n\in\mathcal{N}$ at time $t\in T$. Each vehicle has a target SOC $s_{n,r}$ at the time of leaving the CS. We assume that the vehicle $n$ informs of its leaving time $t_{n,l}$ at the time of plug-in to the CS, e.g., analogous to the way a car is parked at a multi-space parking meter~\cite{multi-space-parking:2011}, so that the CS can confirm $s_n(t_{n,l})=s_{n,r},~\forall n\in\mathcal{N}$. The CS also has the capability to trade its energy with the main grid according to the selling and buying energy rate $e_{g,\text{in}}(t)$ and $e_{g,\text{out}}(t)$ at each $t\in T$.
\begin{figure}[h]
\centering
\includegraphics[width=\columnwidth]{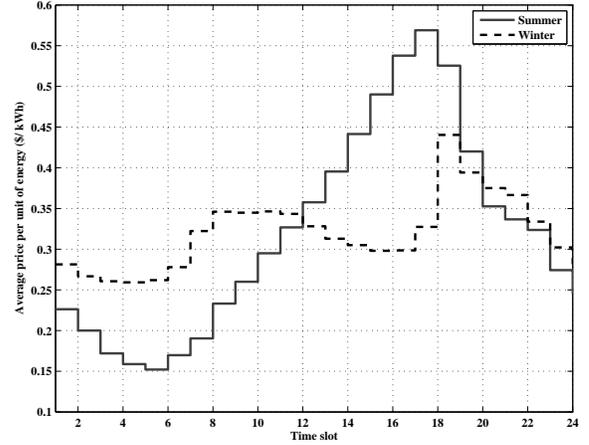}
\caption{Daily electricity wholesale price taken from~\cite{PJM:2013}.} \label{figure:Real-time-price}
\end{figure}

Each energy transaction incurs an associated cost/revenue for the CS in line with the price per unit of energy set by the CS and the grid. In this paper, to determine the price per unit of energy for different energy entities, we have considered the average\footnote{Averaged over a month's data set.} real time electricity sale price shown in Fig.~\ref{figure:Real-time-price} as obtained from PJM, which is a regional transmission organization that coordinates the movement of electricity in all parts of $13$ states and the District of Columbia~\cite{PJM:2013} in the USA. We assume that, on the one hand, the grid sells its energy to the CS at a price $p_{g,\text{out}}(t)$ according to Fig.~\ref{figure:Real-time-price} at different times of the day. On the other hand, the CS adds a mark-up price $\rho$~ to the grid price when it sells its energy to the connected EVs~\cite{Jin-J-TVT:2013}.

To that end, we propose a EV classification in this section. We stress that the per unit price that the CS offers to an EV could be different for different vehicles because of their distinctive energy charging behaviors. We assume that a vehicle can individually decide 1) its time of arriving at and leaving from the CS, 2) the charging rate, within the capacity of the CS, at which the vehicle wants to charge its battery, and 3) whether or not to allow the CS to discharge its battery at any time during its stay at the CS.  To this end, we classify the total number of vehicles in the CS into three categories that are 1) premium vehicles, 2) conservative vehicles, and 3) green vehicles.
\begin{figure}[t]
\centering
\includegraphics[width=\columnwidth]{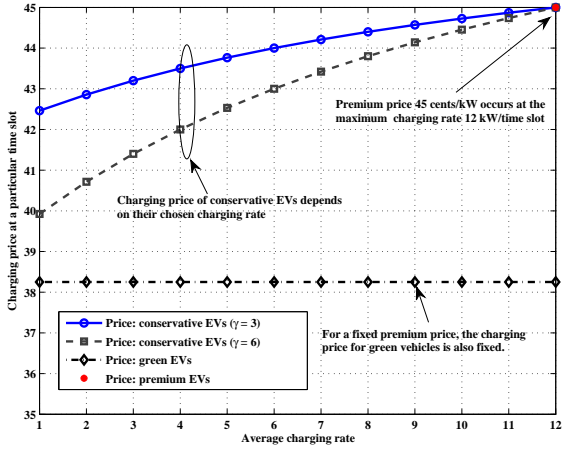}
\caption{Illustration of pricing rate at a particular time slot for different type of EVs where the price per unit of energy for conservative EVs depends on their chosen charging rate.} \label{figure:conservative_price}
\end{figure}
%
\begin{figure}[t]
\centering
\includegraphics[width=\columnwidth]{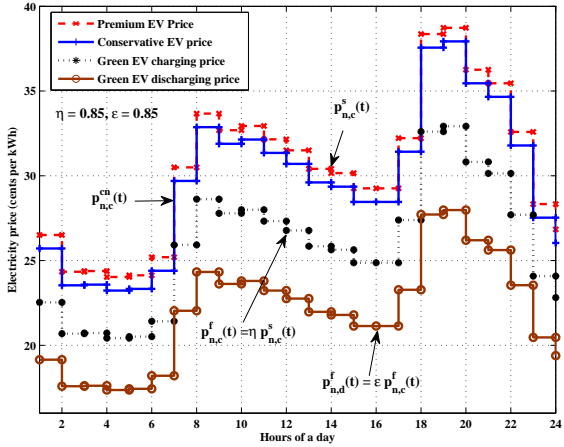}
\caption{Pricing rate of different type of vehicles set by the CS. For conservative EVs, the pricing rate is shown for one particular average charging rate only which can differ for different average charging rates.} \label{figure:pricing_flexible}
\end{figure}
\subsection{Premium vehicles}\label{sec:premiumEV} These type of vehicles are in the set of $\mathcal{N}_s\subset\mathcal{N}$. Each vehicle $n\in\mathcal{N}_s$ charges its battery at the maximum available charging rate, i.e., the premium rate
\begin{eqnarray}
e_{n,c}^s(t) = e_{n,\text{max}},
\label{eqn:premium-charging-rate}
\end{eqnarray}
$\forall n\in\mathcal{N}_s$ and pays the premium price
\begin{eqnarray}
p_{n,c}^s (t)= p_{g,\text{out}}(t)+\rho,
\label{eqn:premium-price}
\end{eqnarray}
$\forall n\in\mathcal{N}_s$ per unit of energy  for charging. Although a premium EV may also want to charge its battery at a rate greater than $e_{n,\text{max}}$, we assume that  the feasible solution is only obtained if this type of vehicle is charged at $e_{n,\text{max}}$, as the CS cannot charge an EV beyond its rated capacity~\cite{Jin-J-TVT:2013}. This type of vehicle does not allow any discharge of their batteries. In the case of early departure from the CS, i.e., if a premium EV leaves the CS before its assigned leaving time $t_{n,l}$, premium users get the best SOC possible for their batteries at the time of leaving.
\subsection{Conservative vehicles}\label{sec:conservativeEV} The set of these type of vehicles is $\mathcal{N}_c\subset\mathcal{N}$ and vehicles in this category charge their batteries at an average rate of
\begin{equation}
e_{n,c}^{cn} = \frac{s_{n,r}-s_{n}(t_{n,a})}{t_{n,l}-t_{n,a}},~\forall n\in\mathcal{N}_c.
\label{eqn:conserv-charge-rate}
\end{equation}
We propose that the CS sets a price
\begin{equation}
p_{n,c}^{cn} (t)= p_{n,c}^s(t) +\gamma - 2\gamma\frac{e_{n,c}^s}{e_{n,c}^s+e_{n,c}^{cn} }
\label{eqn:conserv-price}
\end{equation}
per unit of energy for each conservative vehicle $n\in\mathcal{N}_c$ based on its rate of charging $e_{n,c}^{cn}$. Here, $\gamma$ is a predetermined parameter by the CS. Please note that arrival times, $t_{n,a}$, and leaving times, $t_{n,l}$, are random and hence would be different for each conservative EV  $n\in\mathcal{N}_c$. Consequently, according to \eqref{eqn:conserv-charge-rate}, $e_{n,c}^{cn}$ is different for separate $n\in\mathcal{N}_c$. We also note that $p_{n,c}^{cn} <p_{n,c}^s$ for $e_{n,c}^{cn}>e_{n,c}^s$, which is in fact a reward from the CS to the conservative vehicles for charging the EVs' batteries at a lower rate than the premium. The change of price per unit of energy with the charging rate of a conservative vehicle for different values of $\gamma$ is shown in Fig.~\ref{figure:conservative_price}. Nevertheless, if $e_{n,c}^{cn}\geq e_{n,c}^s$, a conservative vehicle will be treated as a premium EV and its battery will be charged at the maximum charging rate at a cost of premium price $p_{n,c}^{s}(t)$ per unit energy. For conservative vehicles, we further assume that they do not allow the CS to discharge their batteries, and in case of early leaving from the CS, the SOC of their batteries would be lower than their required SOC.
\begin{table*}[t]
\centering
\small
\caption{Summary of the characteristics of proposed different type of EVs.}
\begin{tabular}{|c|c|c|c|}
\hline
& Premium & Conservative & Green\\
\hline\hline
Charging rate & Maximum & $0$ to $e_{n,\text{max}}$, i.e., $e_{n,c}^{cn}(t) = \frac{s_{n,r}-s_{n}(t_{n,a})}{t_{n,l}-t_{n,a}}$& Can be negative\\
\hline
Charging price & Highest price & Depending on charging rate & Lowest\\
\hline
SOC by the time of $t_{n,l}$ & Reach desired state  & Reach desired state & Reach desired state\\
\hline
\shortstack{SOC if leave the CS before \\ the predefined leaving time}  & \shortstack{Highest possible\\ state} & \shortstack{In between initial and\\ desired state} & \shortstack{Can be lower than initial state but definitely\\ higher than a required minimum state}\\
\hline
\end{tabular}
\label{table:comparison}
\end{table*}
 \subsection{Green vehicles}\label{sec:greenEV} These type of vehicles are in set $\mathcal{N}_f\subset\mathcal{N}$, and allow the CS to both charge and discharge their batteries at a rate of $e_{n,c}^f$ and $e_{n,d}^f$ respectively, where $e_{n,c}^f, e_{n,d}^f\in [0,~e_{n,\text{max}}]$. As a reward,
 \begin{enumerate}
 \item they get a discount on their charging price $p_{n,c}^f(t) = p_{n,c}^s(t) \eta$ for charging their batteries, and
 \item they are also paid a price $p_{n,d}^f(t) = p_{n,c}^f(t) \epsilon$ per unit of the energy that they sell to the CS.
 \end{enumerate}
It is important to note that the charging and discharging price of green EVs are connected via $\epsilon$, which is a design parameter determined by the CS. The choice of $\epsilon$ can be affected by the intensity of solar generation, the grid price and the urgency of the CS to buy electricity from the green EVs. However, the choice of $\epsilon$ should be suitable for the green EVs to take part in the energy trading. For example, a value of $\epsilon<<1$ could discourage the EV owners from allowing the CS to discharge their EV batteries if the cost-benefit tradeoff is not attractive. In this context, one possible design value could be $\epsilon = 1$, which makes the charging and discharging price of green EVs the same and price-balanced. In Fig.~\ref{figure:pricing_flexible}, we show an example of charging and discharging price for green EVs, assuming $\epsilon=0.85$, in comparison with a premium price.

In the case of early departure from the CS, depending on the other EVs' demands and PV generation, the SOC of a green vehicle may significantly deviate from its required SOC. For instance, the CS might have just discharged the battery of the EV in the previous time slot to charge other premium and conservative EVs, which have significantly lowered the SOC of that particular green EV. However, we assume that green EVs only allow their batteries to be discharged after a certain minimum amount\footnote{Which could be different for different green EVs.} of SOC $s_{n,f}^\text{min}$ is reached through charging. This SOC is necessary for the green EVs to reach the next nearest charging facility in the case of an emergency early leaving at $t_{n,l}^\text{early}<t_{n,l}$ from the CS. However, if $s_n(t_{n,a})>s_{n,f}^\text{min}$, the EV can allow its battery to be discharged from the initial time slot and $s_n(t_{n,l}^\text{early})$ could be lower than its initial state. Nevertheless, at any time of charging-discharging of all type of EVs the SOC remains above the minimum required amount $s_{n,\text{min}},~\forall n\in\mathcal{N}$, which is necessary for battery safety. We briefly summarize the key characteristics of the proposed three types of EVs in Table~\ref{table:comparison}.

Clearly, $\mathcal{N} = \{\mathcal{N}_s\cup\mathcal{N}_c\cup\mathcal{N}_f\}$. Now, for the energy trading between the CS and the grid, we consider that the grid pays a price $p_{g,\text{in}}(t)$ for buying per unit energy from and charges a price $p_{g,\text{out}}(t)$ to sell per unit of energy to the CS at $t$. In general, $p_{g,\text{in}}(t)<p_{g,\text{out}}(t)$~\cite{McKenna-JIET:2013}. To this end, the CS needs to manage its energy trading with different entities in the energy market, such as the main grid and the three different type of vehicles, at different time slots during $T$. In this context, the proposed classification can easily be implemented in a real-world scenario. For instance, currently vehicles in a pay parking lot need to indicate their arrival  time at and decide on the leaving time from a parking meter before collecting the parking tickets. By incorporating more options at the parking meter, such as whether or not an EV owner wants to discharge his EV's battery and whether or not the EV needs to be charged at a premium rate, the proposed classification of EVs can be established. Now, to capture how the CS can minimize its total cost due to multi-agent energy trading during a day, we design a basic optimization problem for the proposed EV classification scheme in the next section.

\section{MIP Approach with EV Classification}\label{sec:problem-formulation}
In this section, we formulate a basic MIP scheme to capture the benefits of the proposed EV classification scheme in reducing the total cost to the CS for its energy trading with different energy entities throughout a day. We consider an objective function for the CS that it wants to minimize, and design many practical constraints relevant to the studied classification scheme that must be satisfied by the CS, the main grid and each EV during the energy scheduling process by the CS. We note that the studied classification scheme in this paper could also be verified through other existing energy management techniques, e.g., techniques from game theory~\cite{1999Book_Dynamicgame-Basar}, by suitably modifying the CS's objective and constraints for the proposed classification and incentive design. However, we choose MIP over other techniques in this paper motivated by following reasons:
\begin{itemize}
\item The objective function of the CS and the related constraints can be modeled, as we will see in Section~\ref{sec:problem-formulation}, through linear equations and boolean variables\footnote{Boolean variables are required to design some conditional constraints.}. Since, MIP is an optimization technique that is suitable to deal with a linear objective, linear constraints and boolean variables~\cite{Alexander-Schrijver:1998}, we have been motivated to use this technique for this work. Besides, it is relatively simple, as we will see shortly, to capture the impact of the proposed classification of EVs and their charging characteristics on the operational cost of the CS though an MIP approach.\vspace{2mm}
\item MIP has been widely used in smart grid for designing energy management schemes such as in \cite{ChaiBo:SmartGridComm:2014,Hubert-C-ISGT:2011,Parisio-C-SGComm:2011} and \cite{Wille-Haussmann-J-SE:2010}. Therefore, it is a well accepted method for energy trading design within a smart grid paradigm.\vspace{2mm}
\item Various free optimizers, such as Gurobi~\cite{gurobi:2012}, are available online, which can provide optimal solutions to MIP problems in a reasonable time frame.
\end{itemize}
We note that the MIP approach used in this paper provides an optimal offline energy management solution for the proposed classification scheme. This optimal solution can be used as a benchmark in designing more sophisticated online energy management schemes with similar system settings of CS in smart grid.

\hspace{0.1cm}We further note that the proposed MIP necessitates that the prior information on solar generation be known. Nevertheless, this can be extended to a more practical scenario by statistically modeling the future solar generation state, e.g., via a Markov model~\cite{Tushar-ISGT-Europe:2014}, or by using any statistical optimization scheme, e.g., PSO scheme~\cite{Lan-J-NCA:2013,Jian-An-C-ICNSC:2013,Kang-J-TASE:2013}, with incomplete information~\cite{Gunawan-JME:2012} about future solar generation. However, this issue is beyond the scope of this paper\footnote{The motivation for using the basic MIP in this paper is to emphasize the potential benefits that a CS can extract by adopting the proposed classification scheme to reduce its total cost of energy trading.}.
\subsection{Objective function}
The main objective of the CS is assumed to minimize its total cost of energy exchange with different energy entities during a day by suitably choosing the amount of energy that it wants to trade with each vehicle and the grid at each time slot $t\in T$. At any time slot $t$, the cost to the CS depends on the price per unit of energy that it pays/earns through trading its energy with different energy entities, its own PV generation at the roof-top solar panel, the demand of each EV, and the rate at which each green vehicle's battery is discharged by the CS. We assume that the CS can sell energy to/purchase energy from the main grid at any time slot with a price per unit of energy set by the grid. In this regard, the total cost incurred by the CS for a total time duration of $T$ for trading its energy with different energy entities can be expressed as
\begin{eqnarray}
\Gamma &=& \sum_{t=1}^T(\sum_{n = 1}^{N_f} p_{n,d}^f(t) e_{n,d}^f(t) + p_{g,\text{out}}(t) e_{g,\text{out}}(t) -\nonumber \\  &&\sum_{n=1}^{N_f} p_{n,c}^f(t) e_{n,c}^f(t) - \sum_{n=1}^{N_s} p_{n,c}^s(t) e_{n,c}^s(t) - \nonumber\\ && \sum_{n=1}^{N_c} p_{n,c}^{cn}(t) e_{n,c}^{cn}(t) - p_{g,\text{in}}(t) e_{g,\text{in}}(t))\label{eqn:objective}.
\end{eqnarray}
In \eqref{eqn:objective}, all the positive terms are the costs, and all negative terms refer to the revenues to the CS. Thus, the main objective of the CS is to minimize $\Gamma$ by suitably choosing $e_{n,d}^f(t), e_{g,\text{out}}(t), e_{n,c}^f(t)$ and $e_{g,\text{in}}(t)~\forall t\in T$. Here, it is important to note that the CS controls the charging and discharging behavior of green EVs through $\Gamma$, whereby the control over the charging processes of the premium and conservative EVs are accomplished by the CS following the course of actions discussed in Section \ref{sec:premiumEV}, and \eqref{eqn:conserv-charge-rate} and \eqref{eqn:conserv-price} respectively. By contrast, the choice of each EV extends to its choice of being a particular type of EV, i.e., premium, conservative or green, during its period of charging at the CS.

\subsection{Constraints}
While minimizing $\Gamma$, the CS and all connected EVs in the smart grid network need to satisfy a number of constraints so as to ensure that the approach can be applied in a practical environment. Some of these constraints are based on the EV classification scheme that we proposed in this paper.  In the following, we briefly explain the constraints that are assumed to be satisfied by the CS and all three types of EVs in the network during the scheduling of energy trading.
\begin{enumerate}
\item At any time $t$, total supply of energy from the CS to the grid and the connected EVs cannot be more the total energy available during the CS at that time slot. Hence,
\begin{align}
e_{\text{gen}}(t) + e_{n,d}^f(t) +  e_{g,\text{out}}(t) &\geq& e_{n,c}^f(t) + e_{n,c}^s(t)\nonumber \\ &+& e_{n,c}^{cn}(t) + e_{g,\text{in}}(t).\label{eqn:const-1}
\end{align}
\item At the time of plug-in to the grid, the SOC of each vehicle $n$ is equal to $s_n(t_{n,a})$ and during its stay in the CS, i.e., $t_{n,a}<t\leq t_{n,l},~\forall n\in\mathcal{N}$, the SOC is updated based on the SOC from the previous time slot and the amount of energy charging/discharging by the vehicle $n$ at time $t$.
Hence, for $\forall n\in\mathcal{N}$
\begin{equation}
s_n(t) = s_n(t_{n,a}),~\text{if}~t = t_{n,a}.\label{eqn:const-2}
\end{equation}
And, at time $t_{n,a}<t\leq t_{n,l}$
\begin{eqnarray}
s_n(t) =\begin{cases}
s_n(t-1) + \mu e_{n,c}^s(t) ~ \text{if~$n\in\mathcal{N}_s$}\\
s_n(t-1) + \mu e_{n,c}^{cn}(t) ~ \text{if~$n\in\mathcal{N}_c$}\\
s_n(t-1) + \mu e_{n,c}^f(t) - e_{n,d}^f(t),~\text{if} ~n\in\mathcal{N}_f\\
\end{cases}
\label{eqn:const-3}
\end{eqnarray}
where $\mu$ is the efficiency of the battery. Please note that for EVs in the set $\mathcal{N}_f$, i.e., the green EVs, we have an extra term in the equation, which is due to their willingness, unlike the other two types of EVs, to discharge their batteries.
\item At $t_{n,l}$, the SOC of EV $n$ should be equal to at least the amount of SOC they requested and cannot be greater then their capacity.
\begin{align}
s_n (t_{n,l}) \leq b_{n}~\text{and}\nonumber \\
s_n (t_{n,l}) \geq s_{n,r}.
\label{eqn:const-4}
\end{align}
Also, during the time of charging and discharging, the vehicle SOC should not be lower than a minimum value for battery safety as indicated by the battery manufacturer.
\begin{align}
s_n(t)\geq s_{n,\text{min}},~t_{n,a}<t\leq t_{n,l}.
\label{eqn:const-5}
\end{align}
Nonetheless, we note that the green EVs allow their batteries to be discharged by the CS. Hence, in an unlikely event of leaving the CS at $t_{n,l}^\text{early}$, which is before $t_{n,l},~n\in\mathcal{N}_f$, there is a possibility that the EV does not have meaningful SOC in its battery to drive on the road. Hence, we consider that a green EV will allow its battery to be discharged after a minimum required SOC $s_{n,f}^{\text{min}}$  is achieved by the battery so that the EV can travel at least a minimum distance, e.g., to the next available battery exchange station, if it leaves early from the CS. Therefore, in the event of early leaving from the CS
\begin{eqnarray}
s_n(t_{n,l}^\text{early})\geq s_{n,f}^\text{min}~n\in\mathcal{N}_f,~t_{n,l}^\text{early}<t_{n,l}.
\label{eqn:const-early}
\end{eqnarray}
\item For any vehicle $n$, the charging/discharging rate cannot be greater than its rated maximum battery charging capacity. Also, for green vehicles, a vehicle's battery cannot be charged and discharged at the same time. Hence,
\begin{align}
e_{n,c}^s(t), e_{n,c}^{cn}(t)\leq e_{n,\text{max}},~n\in\mathcal{N_s}, \mathcal{N}_c, ~t_{n,a}<t\leq t_{n,l}\nonumber\\
e_{n,c}^f(t)\leq y_{n}(t) e_{n,\text{max}},~n\in\mathcal{N}_f,~t_{n,a}<t\leq t_{n,l}\nonumber\\
e_{n,d}^f(t)\leq (1-y_{n}(t)) e_{n,\text{max}},~n\in\mathcal{N}_f,~t_{n,a}<t\leq t_{n,l},
\label{eqn:const-6}
\end{align}
where, $y_n(t)$ is a binary variable associate with each green vehicle $n\in\mathcal{N}_f$ at time $t$ such that
\begin{align}
y_n(t)=\begin{cases}
1, &\text{if green vehicle $n$ is charging}\\
0, &\text{if green vehicle $n$ is discharging}.
\end{cases}
\nonumber
\end{align}
\item Finally, at any time $t\in T$, the amount of energy flowing to/from the grid should be less than or equal to the grid's maximum power flow capacity. Also, energy cannot flow in to at the same time as it flows out of the grid,
\begin{align}
e_{g,\text{in}}(t)\leq x(t) e_{g,\text{max}}\\
e_{g,\text{out}}(t)\leq (1-x(t)) e_{g,\text{max}},
\label{eqn:const-7}
\end{align}
where
\begin{align}
x(t)=\begin{cases}
1, &\text{if power flows into the grid}\\
0, &\text{if power flows out of the grid}.
\end{cases}
\nonumber
\end{align}
\end{enumerate}
\subsection{Optimizing the scheduling}
\begin{table*}[t!]
\caption{Parameters and variables for MIP}
\centering
\small
\begin{tabular}{l l}
\hline\hline
\textbf{Inputs}\\
\hline
$T\in\mathbb{N}$ & Duration of total scheduling\\
$t\in\{1, 2, \hdots, T\}$ & Time index\\
$N_s, N_c, N_f\in\mathbb{N}$ & Number of premium, conservative and green vehicles\\
$e_\text{gen}(t)\in\mathcal{R}_{+}$\ & Roof-top solar generation at each time slot $t$\\
$s_{n,r}\in\mathcal{R}_{+}$ & Required SOC of each EV at the time of leaving the CS\\
$s_{n,\text{min}}\in\mathcal{R}_{+}$ & Minimum amount of charge at the battery\\
$s_{n,f}^\text{min}\in\mathcal{R}_{+}, n\in\mathcal{N}_f$ & Minimum required SOC before allowing the discharge of the battery\\
$e_{n,c}^s(t),~e_{n,c}^{cn}(t)\in\mathcal{R}_{+},~n\in\mathcal{N}_s,\mathcal{N}_c$ & Charging rate of each premium and conservative vehicle\\
$p_{n,c}^s(t), p_{n,c}^{cn}(t),p_{n,c}^{f}(t)\in\mathcal{R}_{+},~\forall n$ & Charging price per unit of energy for each type of vehicle\\
$p_{n,d}^f(t)\in\mathcal{R}_{+},~n\in\mathcal{N}_f$ & Discharging price per unit of energy for each green vehicle\\
$p_{g,\text{in}}(t), p_{g,\text{out}}(t)\in\mathcal{R}_{+},~\forall n$ & price per unit of energy for buying and selling energy by the grid\\
$e_{n,\text{max}},~e_{g,\text{max}}\in\mathcal{R}_{+}$ & Maximum rate of charging for each vehicle and the grid\\
$b_n\in\mathcal{R}_{+},~n\in\mathcal{N}$ & Capacity of each vehicle's battery\\
$t_{n,a}, t_{n,l}\in\left[1,~T\right],~\forall n\in\mathcal{N}$ & Arrival and leaving time of each vehicle.\\
\hline
\textbf{Variables}\\
\hline
$e_{n,c}^f(t), e_{n,d}^f(t)\in\left[0,~e_{n,\text{max}}\right]$ & Rate of charging of green vehicles. \\
$e_{g,c}(t), e_{g,d}(t) \in\left[0, e_{g,\text{max}}\right]$ & Rate of energy flow into/ out of the main grid.\\
$s_{n}(t)\in\left[s_{n,\text{min}},~b_{n}\right],~n\in\mathcal{N}_f$ & Battery SOC of each vehicle.\\
$x(t)\in\{0,1\}$ & Binary variable to determine the direction of flow.\\
$y_n(t)\in\{0,1\}$ & Binary variable to determine charging/discharging of each green vehicle\\
\hline\hline
\end{tabular}
\label{table:number1}
\end{table*}
After defining the cost function and related constraints, we model the energy trading between the CS, the grid and different EVs in the network as a basic MIP problem. MIP is either the minimization or maximization of a linear function subject to linear constraints where there are some integer variables, e.g., $x(t)$ and $y(t)$ in the proposed case. It is noteworthy that MIP has been used in the literature for various power optimization problems in smart grids, such as in \cite{Hubert-C-ISGT:2011,Parisio-C-SGComm:2011,Wille-Haussmann-J-SE:2010}. We summarize the parameters and variables for the the MIP problem in Table~\ref{table:number1} to optimally minimize the total cost to the CS $\Gamma$. To that end, the optimization problem can be expressed as
\begin{equation}
\min_{\{e_{n,d}^f(t), e_{g,\text{out}}(t),e_{n,c}^f(t),e_{g,\text{in}}(t)\}}\!\!\!\!\!\!\!\!\!\!\!\!\!\Gamma,
\label{eqn:optimization-problem}
\end{equation}
such that constraints~\eqref{eqn:const-1}, \eqref{eqn:const-2}, \eqref{eqn:const-3}, \eqref{eqn:const-4}, \eqref{eqn:const-5}, \eqref{eqn:const-early}, \eqref{eqn:const-6} and \eqref{eqn:const-7} are satisfied. We note that the time of arrival, $t_{n,a}$,  and leaving, $t_{n,l}$, of any vehicle $n\in\mathcal{N}$ are always random and, thus, can be considered as random variables with a suitable probability distribution~\cite{Wayes-J-TSG:2012}. The Gurobi optimizer~\cite{gurobi:2012} is used to solve the optimization problem in  \eqref{eqn:optimization-problem} subject to the related constraints in \eqref{eqn:const-1} to \eqref{eqn:const-7}. We find that the Gurobi optimizer finds the \emph{optimal solution} of the formulated MIP problem \eqref{eqn:optimization-problem}  in a reasonable time-frame while satisfying all the constraints. In the next section, we show the effectiveness of the proposed EV classification scheme via studying the properties and feasibility of the executed optimal scheduling through numerical experiments.

%
\section{Case Study}\label{sec:numerical-simulation}
In this section, we provide extensive simulation results to show the beneficial properties of the proposed EV classification scheme by adopting the scheme with the basic MIP approach studied in the previous section. We show how the mentioned classification can help the CS in reducing its total cost of energy exchange during a day in both summer and winter seasons of the year. In this regard, we consider that public CSs will be needed at places such as on-street parking, at taxi stands, fast-food restaurants and at people's workplaces, and installation of a very large roof-top solar panel in some of these places would be inconvenient. Hence, we consider a small CS consisting of spaces for 24 vehicles spaces. The total time of scheduling, i.e., $7.00$ am to $6.00$ pm, is considered to be divided into $22$ time slots, and each time slot is assumed to have a duration of $30$ minutes~\cite{Anderson-W-EIS:2010}. Unless stated otherwise, it is considered that the number of premium, conservative and green types of vehicles are equal in number at the CS, i.e., eight vehicles of each type are present. The capacity of each vehicle $b_n,~\forall n$ is uniformly randomly chosen from the range $[25, 40]$ kW, which is within the range of typical EV battery capacity~\cite{battery-capcity:2014}. The minimum required SOC at the time of leaving the CS for all EVs is assumed to be $80\%$ of their rated capacity whereas $s_{n,\text{min}},~\forall n$ is considered to be $20\%$ of the battery capacity.  At the time of plug-in to the CS, we consider that the SOC of EVs' batteries are randomly between $20$ to $30\%$ of their capacities. The arrival and leaving time of each vehicle is modeled through a uniformly distributed random variable in the range $[1,~22]$, motivated by \cite{Yifeng-J-TSG:2012,Rashid-J-TRET:2013,Huang-ACC:2012}. It is important to note that the proposed scheme is equally applicable for other choices of arrival and leaving times of EVs as well. Also, if the duration of each time slot changes, e.g., each time slot has 15 minutes duration, the proposed scheme can similarly handle the energy trading if the PV output is measured over every time slot. This can be achieved by sampling and recording the solar irradiance data for every 15 minutes  duration and then calculating the PV output via \eqref{eqn:energy-basic} and \eqref{eqn:solar-output}.

It is assumed that each roof-top solar panel has a dimension of $1.926\times1.014$~square meters (m)~\cite{SolarPanel} and that the required floor area for each vehicle space is $2.4\times 4.8$ m$^2$~\cite{Parking:2005}. Consequently for $N$ vehicles, the total number of solar panels on the rooftop is $K = \frac{2.4\times 4.8}{1.926\times1.014} N$. The value of $I_\text{light}(t)$ at each time slot $t\in T$ is considered from the set of solar data (measured at the Australian National University (ANU) in Canberra, Australia), which is averaged over a month's data for the considered time slots. The selling price $p_{n,\text{out}}(t)$ per unit of energy by the grid is considered as the same as shown in Fig.~\ref{figure:Real-time-price}, where as the buying price $p_{n,\text{in}}(t)$ is assumed to be $p_{n,\text{out}}(t) - \omega,~\omega>0$. As for the pricing rate for charging the batteries of connected premium and conservative vehicles, $p_{n,c}^s(t) = p_{n,\text{out}}(t)+\rho,~n\in\mathcal{N}_s$, and $p_{n,c}^{cn}(t),~n\in\mathcal{N}_c$ are respectively chosen by the CS according to \eqref{eqn:conserv-price}. For each green vehicle $n\in\mathcal{N}_f$, $p_{n,c}^f(t)$ and $p_{n,d}^f(t)$ are chosen as $p_{n,c}^s(t)\eta$ and $p_{n,c}^f(t)\epsilon$ respectively. In the following experiments, we have chosen $\omega = 2, \rho = 5, \eta = 0.75$ and $\epsilon = 0.85$. We note that the choice $\epsilon$ can significantly affect the convergence of the proposed scheme to an optimal solution as explained in Section~\ref{sec:greenEV}. Essentially, if it is significantly lower than $1$, e.g., $\epsilon<0.75$, the MIP does not converge to an optimal solution within a reasonable time frame. This property, in fact, can be used by the CS to doubly confirm a suitable value of $\epsilon$ to run the MIP in order to facilitate the energy trading by the green EVs. For instance, if $\epsilon\geq 0.79$, the proposed MIP converges to the optimal solution for all suitable combinations of other system parameters within a reasonable time-frame (less than 5 seconds), and thus the computational complexity of the scheme reduces significantly. Unless stated otherwise, the solar data, used in the derivation of the results of each figure, are data for summer and the number of different type of EVs are considered to be the same in the CS\footnote{The values of $\omega, \rho, \eta$ and $\epsilon$ are chosen to make the studied cases consistent with the pricing schemes explained in Section~\ref{sec:system-model}. However, the parameters may have different values at different places, for different trading schemes and under different weather conditions. For example, the technique can be easily evaluated for a price-balanced pricing scheme for green EVs by setting the value of $\epsilon$ equal to 1.}.
\begin{figure}[t!]
\centering
\includegraphics[width=\columnwidth]{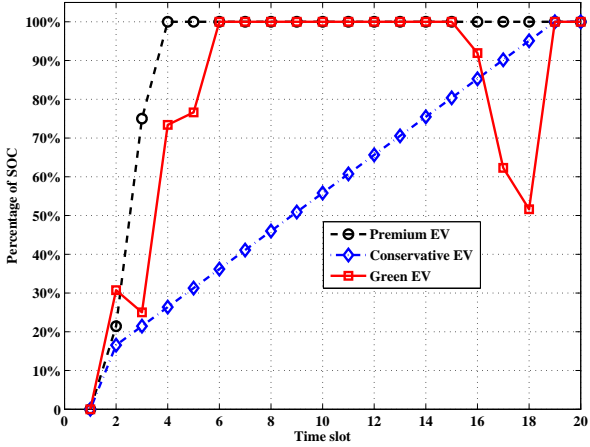}
\caption{Charging discharging behavior of each type of EV.}
\label{fig:fig-a}
\end{figure}
\begin{figure}
\centering
\includegraphics[width=\columnwidth]{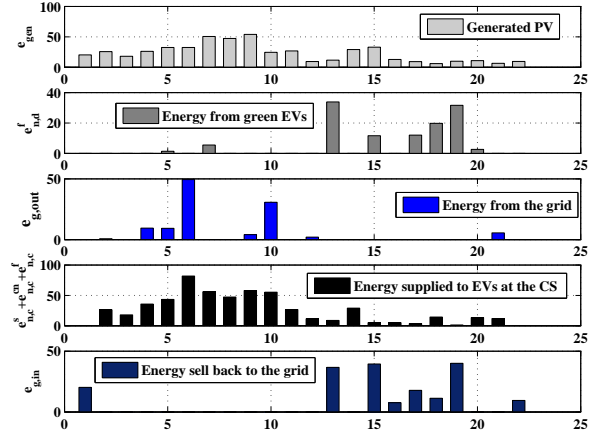}
\caption{Charging and discharging behavior of the grid and CS.}
\label{fig:fig-b}
\end{figure}

Now, to observe the performance of the proposed scheme, we show some related data to demonstrate the charging/discharging pattern of each different type of EV, and of the CS and the grid in Fig.~\ref{fig:fig-a} and Fig.~\ref{fig:fig-b} respectively. First, in Fig.~\ref{fig:fig-a}, the charging and discharging behavior of premium, conservative and green EVs are shown for their arrival ($t_{n,a} = 2$) and leaving times ($t_{n,l} = 20$), which we keep similar for all three types of EVs for this particular case. The SOC of each type of EV is normalized to its target SOC at the time of leaving the CS. In general, as can be see from Fig.~\ref{fig:fig-a}, all three different types of EVs reach their target SOC levels before leaving the CS. However, the charging patterns are different for different EVs. As the scheme is designed, and also can be seen from Fig~\ref{fig:fig-a}, a premium EV is charged at the highest rate and reach its target 100\% SOC faster than other EVs in the CS; the charging of a conservative EV is accomplished at an average rate determined by the EV's initial and target SOC and the arrival and leaving time; and finally, for green EV, the charging is taking place during time slots $4, 5$ and $6$. The battery of the EV is used as energy storage by the CS from  time slots $7$ to $14$ and then discharged at time slots $15, 16$ and $17$ respectively. Finally, the SOC of the green EV reaches its target $100\%$ through charging at time slot $18$. From Fig.~\ref{fig:fig-a}, we further note that none other than the green EV allows the CS to discharge its battery as discussed in the proposed scheme.

In Fig. \ref{fig:fig-b}, we show data on energy trading by the CS and the grid along with the generation of solar energy at different time slots of a typical day. As designed, we note that the grid does not sell ($e_{g,\text{out}}$) and buy ($e_{g,\text{in}}$) energy concurrently in the same time slot. In time slots $4, 5$ and $6$, there is not enough energy generated from the PV cells to satisfy the charge requests of the charging EVs; hence the CS draws down some charge from the green EVs acting as distributed storage to supplement. In contrast, in time slots such as $15$ and $19$, the CS has excess generation that needs to be stored, and hence pushes this excess generation to the connected green EVs. It is also important to note that, in any time slot, the total sum of energy received by the CS, i.e., from its PV ($e_\text{gen}$), green EVs ($e_{n,d}^f$) and the grid ($e_{g,\text{out}}$) at least equals the electricity that the CS supplies to  all the charging EVs in the CS ($e_{n,c}^f + e_{n,c}^{cn}+e_{n,c}^f$) and to the grid ($e_{g,\text{in}}$), as stated in \eqref{eqn:const-1}; this is evident throughout Fig.~\ref{fig:fig-b}.
\begin{figure}[t]
\centering
\includegraphics[width=\columnwidth]{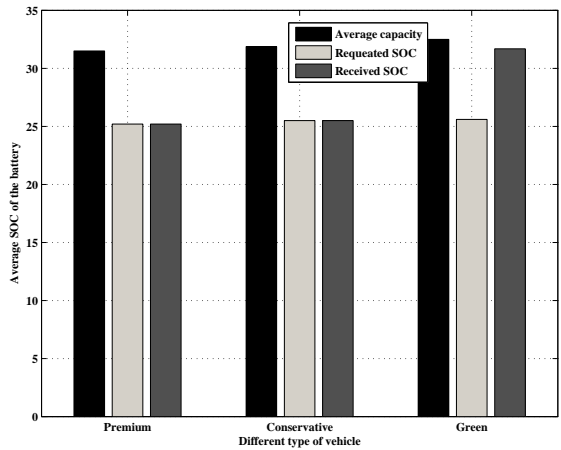}
\caption{Average SOC of different types of vehicles at the time of leaving the CS.} \label{figure:Mean_SOC_of_Each_Vehicle}
\end{figure}

Now, after a brief overview of the energy trading behavior of different energy entities, in Fig.~\ref{figure:Mean_SOC_of_Each_Vehicle} we show the average SOC per EV, which is averaged over eight vehicles for each different type of EV, at the time of leaving the CS on a typical summer's day. As can be seen from the figure, all vehicles have their minimum required SOC when they leave. Noticeably, no type of vehicle other than a green type is capable of gaining more than their minimum requirement. This is due to the way that the problem is formulated in the paper. In fact, the CS can not discharge the battery, or set the charging rate, for both premium and conservative vehicles. Consequently, these vehicles are only being charged up to the amount that they request, and not more than that. However, for green vehicles,  some EVs may have received more energy than their minimum requested amount by the time they leave. This solely depends on the leaving time, the demand of other EVs in the CS, and the solar electricity generation from the rooftop solar panel. Essentially, if the CS has more energy than it needs for a particular time slot, it can reserve this energy in the batteries of green vehicles. As a result, on average, the SOC at the time of leaving could be higher than the minimum SOC requirement for the green vehicles as demonstrated in Fig.~\ref{figure:Mean_SOC_of_Each_Vehicle}.
\begin{figure}[t]
\centering
\includegraphics[width=\columnwidth]{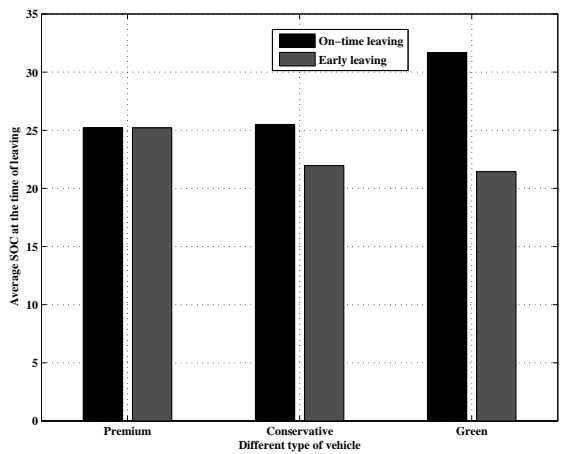}
\caption{Average SOC of different type of vehicles in case of leaving two time slots earlier than projected leaving time.} \label{figure:Early_leavingSOC}
\end{figure}

Nonetheless, a very interesting insight is obtained by observation of the case when a vehicle wants to leave the CS earlier than its projected time, i.e., the leaving time that it originally informs to the CS. In Fig.~\ref{figure:Early_leavingSOC}, we plot the average received SOC per EV (averaged over eight EVs of each type), at its actual leaving time when that is two time slots earlier than the projected time. As we can see from the figure, the average SOC of the premium vehicle is very close to that of the case when it leaves on time. This is due to the fact that they are always being charged at the maximum charging rate, and consequently, their batteries reach the required SOC very quickly. As a result, if the EV leaves early under any circumstances, its battery is equal to (in some instances, very close to) their requested SOC. On the other hand, conservative vehicles achieve an average SOC noticeably lower than their required  amount in the case of early leaving. These type of vehicles charge their battery at an average charging rate calculated via ~\eqref{eqn:conserv-charge-rate}. Hence, if any vehicle leaves earlier than they originally stated, they are not able to charge their battery to their required SOC level. Interestingly, the maximum deviation of the SOC level compared to the case where each vehicle leaves at the time they originally stated is observed in the case of green vehicles. Although, the average SOC per green vehicle is greater than their minimum requirement in the case of leaving on time, leaving two time slots early degrades the average SOC level significantly. This is because the CS charges and discharges the batteries of green vehicles in order to minimize its total cost. Hence, the SOC of a green vehicle could be very low at two time slots before its leaving time. For instance, the CS can use the green EV's battery energy to charge some other vehicles or to sell it to the grid when it is two time slots or more before its stated departure time, and again recharge its battery at a higher charging rate in the next two time slots. As a result, the average SOC is much lower for green vehicles when they leave early, as shown in Fig.~\ref{figure:Early_leavingSOC}. Therefore, based on the observation of SOC of EVs in the case of early leaving, we can conclude that for EVs who choose to behave as green EVs in the CS, they should not leave the CS at any time before their projected leaving time in order to reduce the risk of having a significantly lower SOC than required. However, the SOC level of any green EV is always maintained at least to $s_{n,f}^\text{min}$, even if they leave early, according to \eqref{eqn:const-early}.%
\begin{figure}[t]
\centering
\includegraphics[width=\columnwidth]{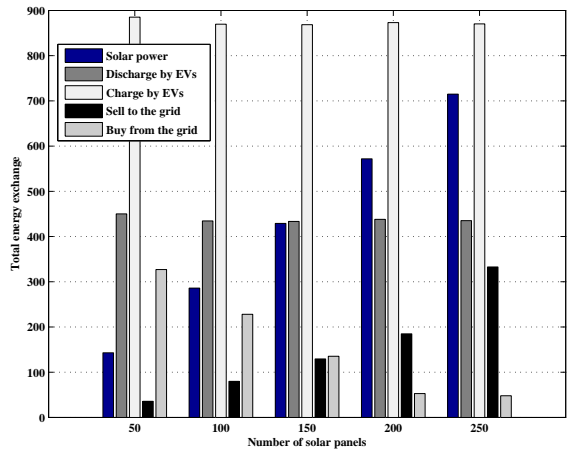}
\caption{Change of the flow of total energy through different entities with the change in the number of solar panels on the roof-top.} \label{figure:DIfferent_energy_flow}
\end{figure}

In Fig.~\ref{figure:DIfferent_energy_flow}, we show how the flow of the total amount of energy via different energy entities is affected as the number of roof-top solar panels increases in the CS. We assume the same EV traffic conditions in all cases. Due to the same number of vehicles with the same demands for energy, the total amount of energy flowing into the CS is always identical for any given number of panels. Interestingly, the amount of power that the CS buys from the green EVs also does not change noticeably with a change in the number of solar panels. This is mainly because the CS pays at the lowest rate to buy power from green EVs compared to any other energy entities. Hence, even if the generation from solar energy is higher, the CS wants to buy the same amount of energy from the green vehicles so as to sell more energy to the grid to make more revenue, thus further lowering the total cost. Nonetheless, a noticeable change is observed in the amount of energy both bought from the grid, and sold to the grid, by the CS with the change in the number of solar panels. As shown in Fig.~\ref{figure:DIfferent_energy_flow}, as the number of solar panels increases the amount of energy that the CS buys from the grid decreases as the CS's own production of energy increases. Further, since the CS is interested in minimizing its costs (or, maximizing its revenue), it increases its amount of energy for sale to the grid with an increase in the number of solar panels. However, we have not considered the cost/inconvenience of setting up such a huge CS, which might affect the implementation of this scenario. The demonstrated performance may further vary for different environmental conditions and also for different vehicle traffic conditions.
\begin{figure}[t]
\centering
\includegraphics[width=\columnwidth]{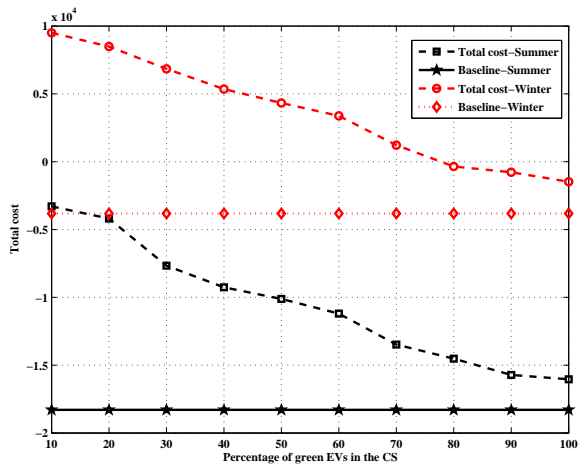}
\caption{Demonstration of the change of total cost to the CS as the percentage of green EVs in the CS. The performance is compared with the case when all EVs that are connected to the CS are green and thus the CS uses MIP to decide on the charging and discharging of each of the EVs in the system. Thus, for both summer and winter seasons, the figure demonstrates a performance comparison between the proposed scheme and an MIP approach.} \label{figure:cost_summe_winter}
\end{figure}

To observe the effect of penetration of green EVs on the total cost to the CS, we run a simulation for a different percentage of green EVs and plot the relevant total cost to the CS in Fig.~\ref{figure:cost_summe_winter}. We consider the average solar generation data for both summer and winter seasons in this case. Here, two important effects are noteworthy: 1) the total cost incurred by the CS decreases as the \emph{percentage of green} vehicles in the CS \emph{increases}, and 2) as the solar generation at the CS increases
, the total cost incurred by the CS decreases. This is due to the fact that the solar generation is significantly higher in summer than in winter, as shown in Fig.~\ref{figure:solar-data}, for example. More electricity is generated from more received solar energy, which enables the CS to sell more to the grid after meeting the demand of connected EVs and, consequently, a lower cost is incurred to the CS during summer than in winter.

On the other hand, an increased percentage of green vehicles enables the CS to buy more energy from these green EVs instead of buying energy from the grid at a higher price. As a consequence, the total cost to the CS reduces even further. To quantify the performance improvement by incorporating green vehicles, we compare this result with a baseline approach. The baseline approach is modeled assuming that all EVs in the CS are green and that there is a \emph{fixed contract} price~\cite{pricing-scheme:2012} for all energy trading between different entities, e.g., energy trading between CS and the grid. This baseline approach is considered as the benchmark that the proposed scheme targets. As shown in Fig.~\ref{figure:cost_summe_winter}, an increase in the penetration of green vehicles in the CS enables the total cost, in both summer and winter seasons, to converge towards the baseline. It is important to note that the cost can become negative as the percentage of green EVs increases. This is because of the way we model our cost function. Negative cost refers to the fact that the CS is in fact receiving net revenue from selling its energy. As seen in the figure, with an increase in penetration of green vehicles from $10\%$ to $100\%$, the difference between the total costs with respect to the baseline approach drops down by $85\%$ for summer, and by $82\%$ for winter.

\begin{figure}[t]
\centering
\includegraphics[width=\columnwidth]{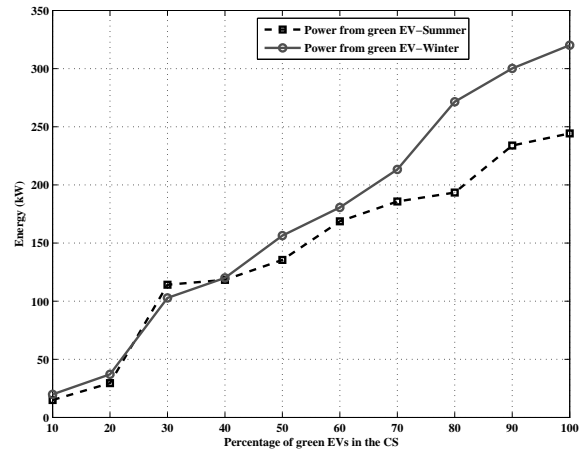}
\caption{Change in total energy that green EVs contribute to the CS as the percentage of green EVs increases.} \label{figure:Flow_of_renewable}
\end{figure}

\begin{table*}[!t]
\centering
\small
\caption{Average cost (in cents) to charge battery of each different type of EV during a day in summer and winter season.}
\begin{tabular}{|c|c|c|}
\hline
 & Summer & Winter \\
\hline\hline
Cost of Premium EV & 677.23 & 735.23\\
\hline
Cost of Conservative EV & 640.20 & 648.54\\
\hline
Cost of Green EV & 354.10 & 371.54\\
\hline
Cost reduction of green EVs compared to premium &47.7\% & 49.4\%\\
\hline
Cost reduction of green EVs compared to conservative &44.6\%&42.71\%\\
\hline
\end{tabular}
\label{table:cost-avg}
\end{table*}

Whereas increasing the number of green vehicles can considerably reduce the total cost to the CS in both summer and winter, it would also be very interesting to see how the green EVs contribute to this reduction in cost. In this regard, we show the change in the total energy that the green vehicles contribute in reducing total cost as the relative percentage of green EVs in the CS changes in Fig.~\ref{figure:Flow_of_renewable}. In winter, solar generation is lower and hence the CS needs to buy a larger fraction of energy from the main grid so as to meet the demand of its connected EVs.  However, as the percentage of green vehicles increases in the system, the CS intends to buy more energy from the green EVs instead of from the main grid so as to reduce total cost. Hence, the cost goes down as explained in Fig.~\ref{figure:cost_summe_winter}. On the other hand, solar generation is higher in summer. Hence, the dependency of the CS on green vehicles is lower in summer than in winter. Therefore, as can be seen from Fig.~\ref{figure:Flow_of_renewable}, the amount of energy sold by green vehicles to the CS is lower in summer than in winter. Nonetheless, due to the presence of ample solar power in summer compared to winter, the reduction in cost is similar in both seasons. Furthermore, it is important to note that an increase in the  number of green EVs and generated solar energy can considerably affect the power quality, e.g., by harmonic distortion and voltage flicker of the CS. One potential way to improve and stabilize the power quality of such a solar energy system is to incorporate intelligent techniques, including artificial neural networks, genetic algorithms and particle swarm optimization~\cite{Alexander :2014}, within the system. Nevertheless, this is beyond the scope of this work, and therefore is not investigated here.

Finally, we show how the proposed classification can help the green EVs to reduce their average cost of a day during summer and winter seasons. We assume that the number of different types of EVs are equal, i.e., eight EVs for each type,  in the CS. To that end, the average cost that is incurred to an EV during summer and winter is shown in Table~\ref{table:cost-avg}.  As can be seen from the table, in both seasons the average total cost to a green EV is considerably lower than for both the premium and conservative vehicles. For example, the cost reduction for a green EV in summer is around $44.6\%$ and $47.7\%$ compared to a conservative and premium vehicle respectively. The similar reduction in cost is also found in winter. Hence, it can be concluded that the proposed classification not only helps the CS to reduce its total cost of energy purchase but also assists EVs to reduce their average total cost if they choose to act as green while connecting to the CS.

\section{Conclusion}\label{sec:conclusion}
In this paper, a classification scheme of electric vehicles (EVs)  has been studied that can assist a photovoltaic (PV) driven charging station (CS) to reduce its total cost of energy trading with different energy entities in a smart grid network during a day. In modeling the classification, the fact that all EV owners might not allow the discharging of the batteries of their EVs is considered, and a novel classification of vehicles has been proposed. Based on the charging behavior the connected vehicles to the CS have been categorized in to three kinds: 1) premium, 2) conservative, and 3) green. The rules of charging for premium, conservative and green EVs have been designed along with the energy discharging rules exclusively for green vehicles. The pricing scheme to charge different EVs based on their charging/discharging behavior has also been discussed. To test the capability of the proposed constraints related to the CS and the grid, using real solar and pricing data, it has been shown that the introduction of green vehicles can significantly reduce the total cost to the CS. The results have been compared with a baseline approach, and it has been shown that, as the number of green EVs increases in the system, the total cost tends to converge more towards the baseline. Furthermore, as the percentage of green vehicles in the CS increases from $10\%$ to $100\%$, the difference for energy trading, between the total cost of the proposed scheme and the baseline approach, has been quantified as reducing by $85\%$ in summer, and by $82\%$ in winter. The benefits to EVs of being green are also demonstrated in terms of reduction in their average total cost of a day charging during summer and winter.

The proposed scheme can be extended and improved in various aspects. The system model can be extended to incorporate a penalty for vehicles who stay after their specified leaving time in the CS, and also to investigate the effect of different choice of arrival and leaving time design models on the cost to the CS. Another interesting extension would be to develop a real-time algorithm, and associated analysis, to capture energy management without the knowledge of future car arrival and leaving times, and uncertainty in both generation from renewables and price. Further, the analysis of power quality as the number of green EVs and the solar energy generation increases in the system is another interesting future research extension.

%
\begin{IEEEbiography}[{\includegraphics[width=1in,height=1.25in,clip,keepaspectratio]{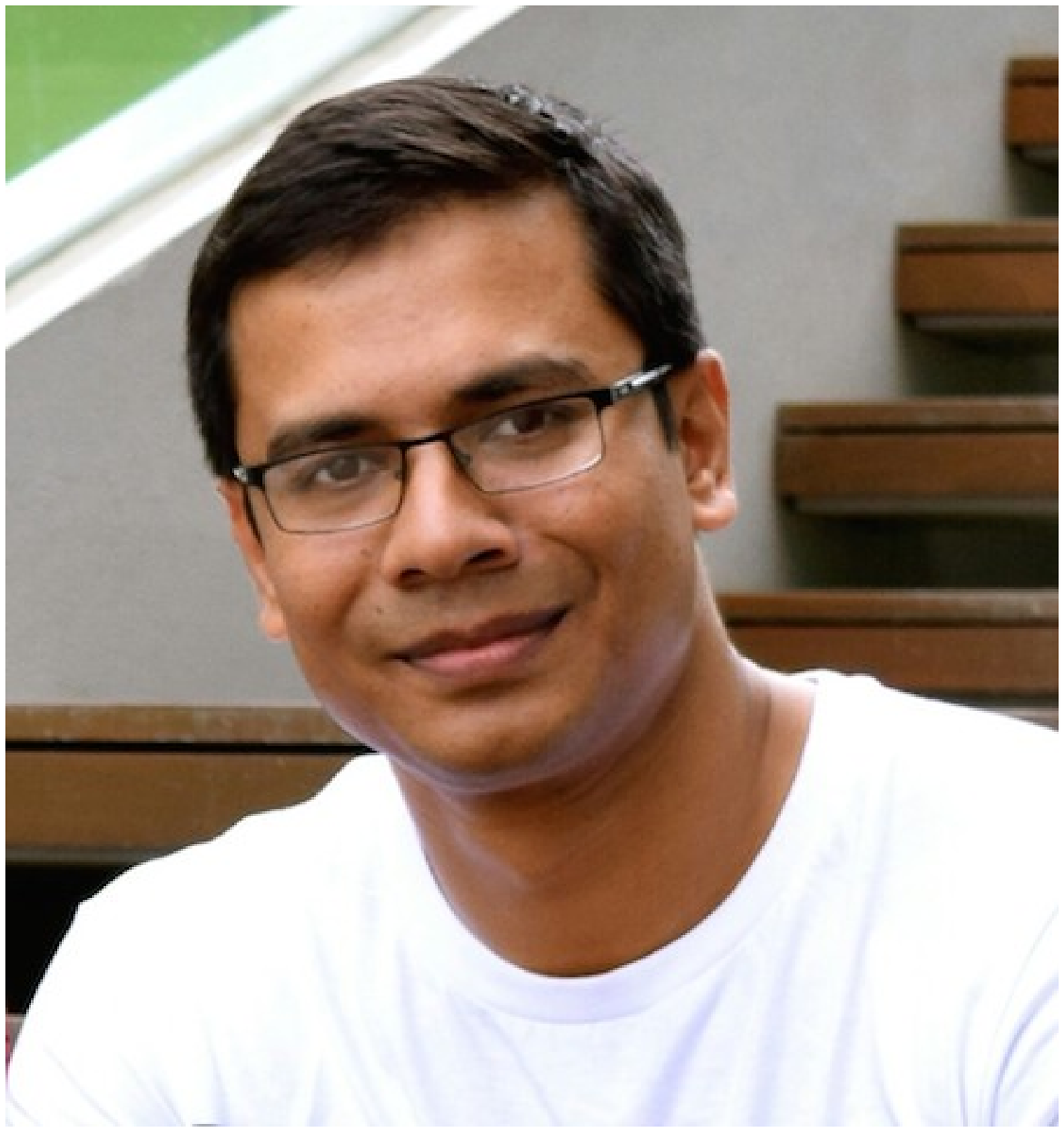}}]
{Wayes Tushar}(S'06, M'13)  received the B.Sc. degree in Electrical and Electronic Engineering from Bangladesh University of Engineering and Technology (BUET), Bangladesh, in 2007 and the Ph.D. degree in Engineering from the Australian National University (ANU), Australia in 2013. Currently, he is a Research Scientist at SUTD-MIT International Design Center in Singapore University of Technology and Design (SUTD), Singapore. Prior joining SUTD, he was a visiting researcher at National ICT Australia (NICTA) in ACT, Australia. He was also a visiting student research collaborator in the School of Engineering and Applied Science at Princeton University, NJ, USA during summer 2011. His research interest includes signal processing for distributed networks, game theory and energy management for smart grids. He is the recipient of two best paper awards, both as the first author, in Australian Communications Theory Workshop (AusCTW), 2012 and IEEE International Conference on Communications (ICC), 2013.
\end{IEEEbiography}

\begin{IEEEbiography}[{\includegraphics[width=1in,height=1.25in,clip,keepaspectratio]{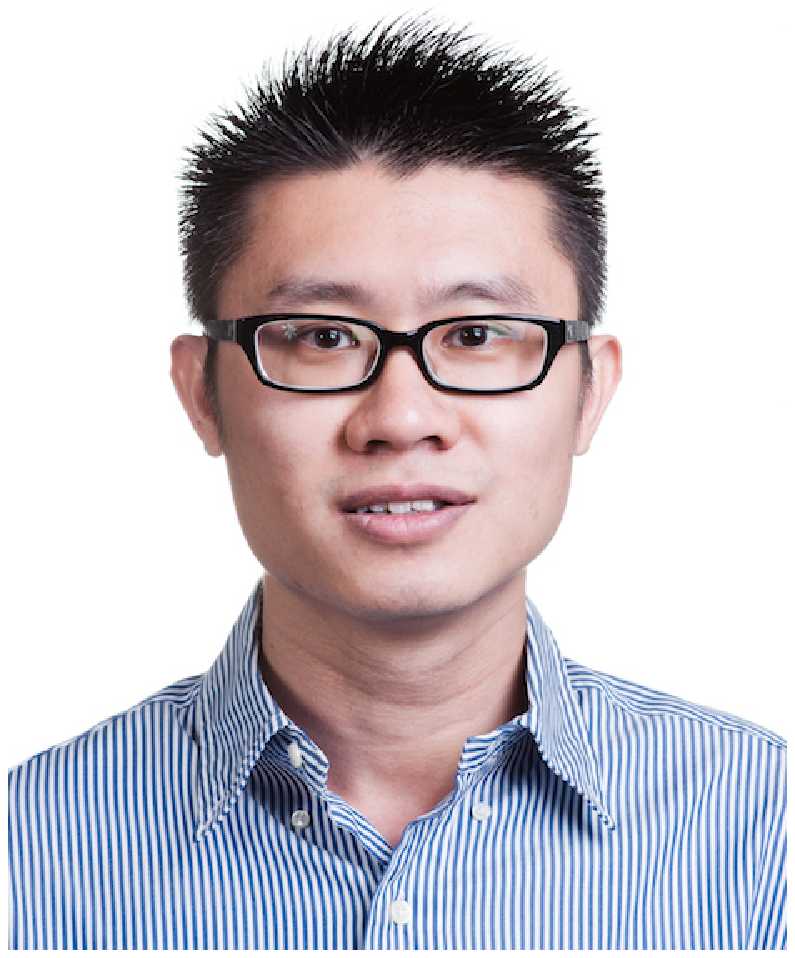}}]
{Chau Yuen} received the B. Eng and PhD degree from Nanyang Technological
University, Singapore in 2000 and 2004 respectively. Dr Yuen was a Post
Doc Fellow in Lucent Technologies Bell Labs, Murray Hill during 2005.
He was a Visiting Assistant Professor of Hong Kong Polytechnic University
in 2008. During the period of 2006 2010, he worked at the Institute for
Infocomm Research (Singapore) as a Senior Research Engineer. He joined
Singapore University of Technology and Design as an assistant professor
from June 2010. He serves as an Associate Editor for IEEE Transactions on
Vehicular Technology. On 2012, he received IEEE Asia-PaciÞc Outstanding
Young Researcher Award.
\end{IEEEbiography}

\begin{IEEEbiography}[{\includegraphics[width=1in,height=1.25in,clip,keepaspectratio]{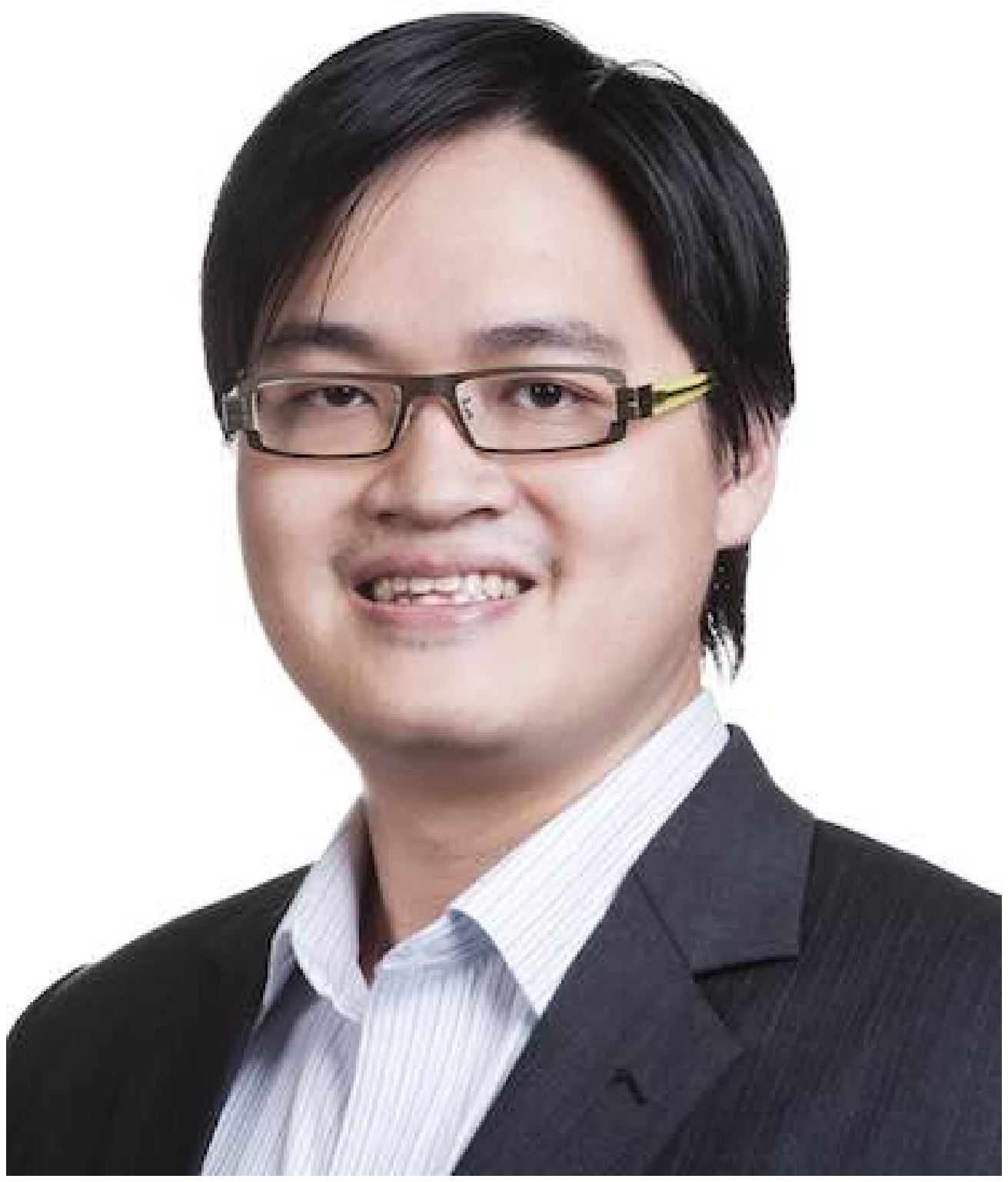}}]
{Shisheng Huang} received his B.Eng. from the National University of Singapore in 2006 and subsequently, his Ph.D. from Purdue University, West Lafayette, IN in 2011. Currently, he is with the Operations Research Unit of the Ministry of Home Affairs, Singapore. His research focuses on systems analysis using simulation and modelling approaches; with applications in energy systems, transportation systems and the smart grid.
\end{IEEEbiography}

\begin{IEEEbiography}[{\includegraphics[width=1in,height=1.25in,clip,keepaspectratio]{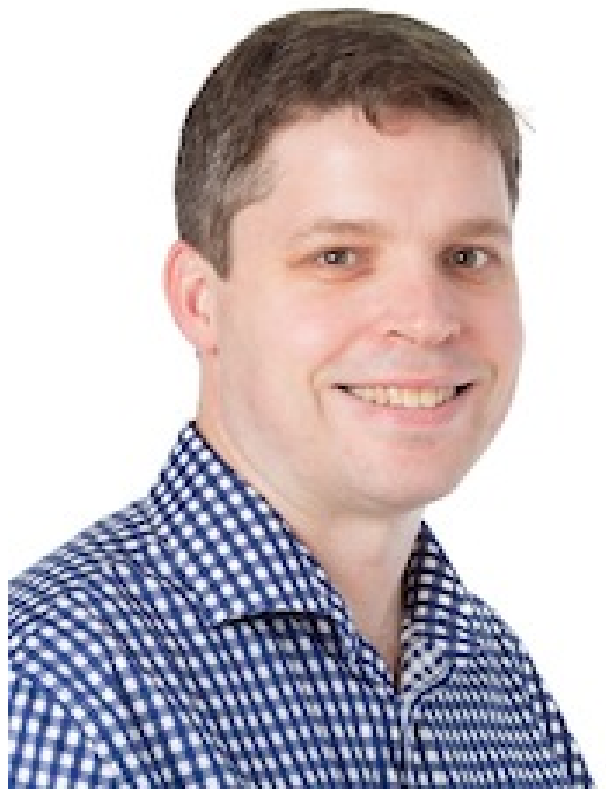}}]
{David Smith}(S'01, M'04)  is a Senior Researcher at National ICT Australia (NICTA) and is an adjunct Fellow with the Australian National University (ANU), and has been with NICTA and the ANU since 2004. He received a B.E. degree in Electrical Engineering from the University of N.S.W. Australia in 1997. He obtained an M.E. (research) degree in 2001, and a Ph.D. degree in 2004, both from the University of Technology, Sydney (UTS) in Telecommunications Engineering. His research interests are in wireless body area networks; game theory for distributed networks; mesh networks; disaster tolerant networks; radio propagation; MIMO wireless systems; space-time coding; antenna design; and also in distributed optimization for smart grid. He has also had a variety of industry experience in electrical and telecommunications engineering. He has published over 80 technical refereed papers and made various contributions to IEEE standardization activity; and has received four conference best paper awards.
\end{IEEEbiography}

\begin{IEEEbiography}[{\includegraphics[width=1in,height=1.25in,clip,keepaspectratio]{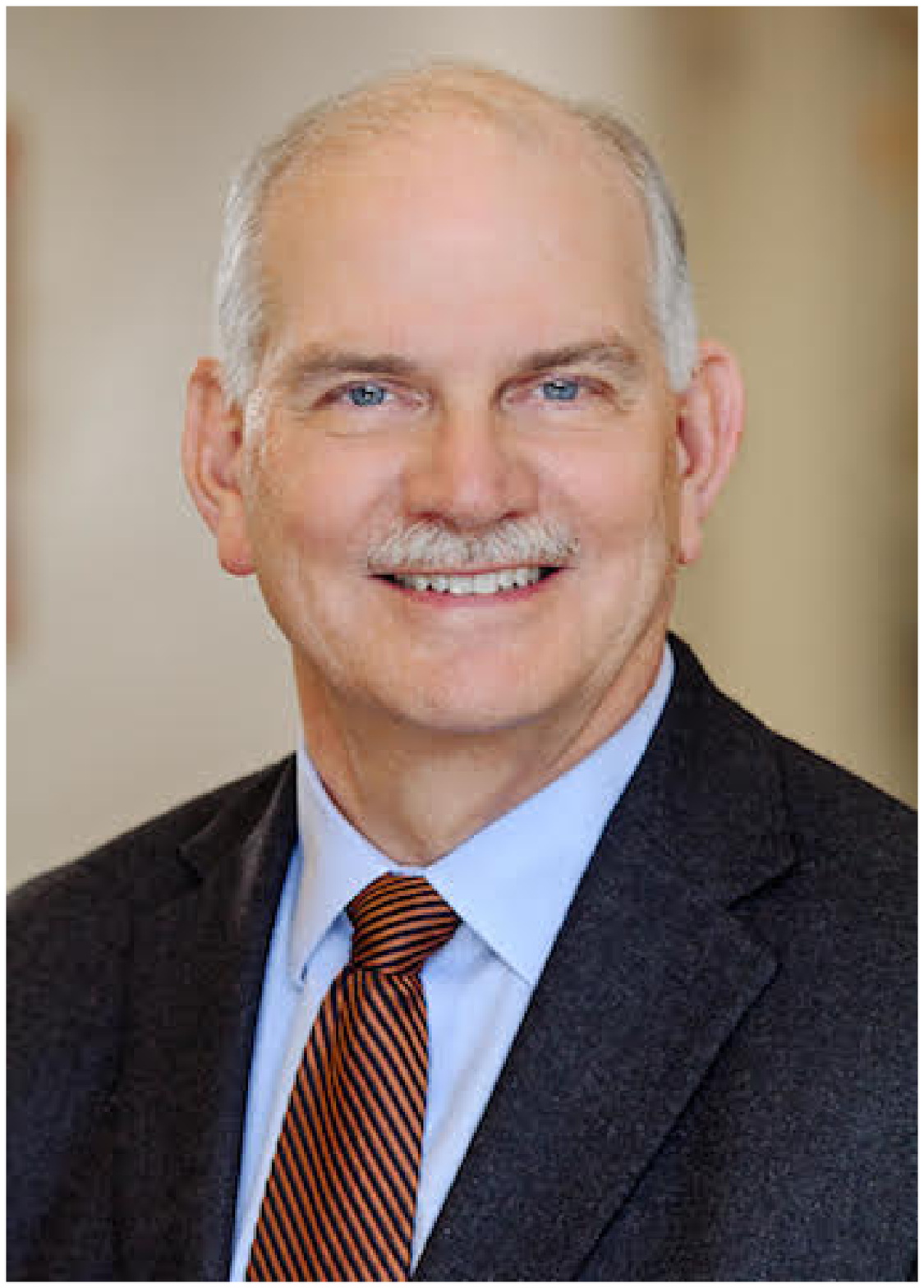}}]
{H. Vincent Poor} (S'72, M'77, SM'82, F'87) received the Ph.D. degree in EECS from Princeton University in 1977.  From 1977 until 1990, he was on the faculty of the University of Illinois at Urbana-Champaign. Since 1990 he has been on the faculty at Princeton, where he is the Michael Henry Strater University Professor of Electrical Engineering and Dean of the School of Engineering and Applied Science. Dr. Poor's research interests are in the areas of stochastic analysis, statistical signal processing, and information theory, and their applications in wireless networks and related fields such as social networks and smart grid. Among his publications in these areas is the recent book \emph{Mechanisms and Games for Dynamic Spectrum Allocation} (Cambridge University Press, 2014).

Dr. Poor is a member of the National Academy of Engineering and the National Academy of Sciences, and a foreign member of Academia Europaea and the Royal Society. He is also a fellow of the American Academy of Arts and Sciences, the Royal Academy of Engineering (U.K), and the Royal Society of Edinburgh. In 1990, he served as President of the IEEE Information Theory Society, and in 2004-07 he served as the Editor-in-Chief of the IEEE TRANSACTIONS ON INFORMATION THEORY. He received a Guggenheim Fellowship in 2002 and the IEEE Education Medal in 2005. Recent recognition of his work includes the 2014 URSI Booker Gold Medal, and honorary doctorates from Aalborg University, Aalto University, HKUST and the University of Edinburgh. 
.
\end{IEEEbiography}
\end{document}